\documentclass[twocolumn,showpacs]{revtex4}
\begin{document}
\title{ On Conformally Flat Stationary Axisymmetric Spacetimes}
\author{Alberto A. Garc\'{\i}a and  Cuauhtemoc Campuzano\\
Departamento~de~F\'{\i}sica.\\~Centro~de~Investigaci\'on~y~de~Estudios~Avanzados~del~IPN.\\
Apdo. Postal 14-740, 07000 M\'exico DF, MEXICO}
\email{aagarcia@fis.cinvestav.mx, ccvargas@fis.cinvestav.mx }
\begin{abstract}
It is shown that within conformally flat stationary axisymmetric
spacetimes, besides of the static family, there exists a new class
of metrics, which is always stationary and axisymmetric. All these
spacetimes, the static and the stationary ones, are endowed with
an arbitrary function depending on the two non--Killingian
coordinates. The explicit form of this function can be determined
once the coupled matter, i.e., the energy--momentum tensor is
given. One might hope possible extensions of this result to black
holes on two--branes in four dimensions.
\end{abstract}


\maketitle
\date{\today}

\section{Introduction}
In 1976, Collinson~\cite{col} formulated the following theo\-rem:
{\it Every conformally flat stationary axisymmetric space--time is
necessarely static. If the source of such space--time is a perfect
fluid, then the space--time metric can be reduced to the usual
Schwarzschild interior metric}. The main goal of present work is
to establish the existence of a new class of intrinsically
stationary axi\-sym\-me\-tric spacetimes, which, at the same time,
are conformally flat. Hence, the Collinson theorem fails to be
true in its first statement. In what follows the complex
coefficients asso\-cia\-ted to the conformal Weyl tensor for the
general stationary axisymmetric metric are given. Next, in Sec.
\ref{ii} the general integrals for conformally flat spaces are
presented. Sec. \ref{iii} deals with conformally flat locally
static spaces. As a by--product, we demonstrate here that one can
isolate spacetimes of the form $R \times
BTZ$~\cite{emparan,klgar}; the results of ~\cite{emparan} suggest
the possibility of constructing four dimensional black holes bound
to two--branes. In Sec. \ref{iv} the general expression of
conformally flat stationary axi\-sym\-me\-tric metrics is given.
Finally, some concluding remarks are stated.

The starting point in our study is the stationary a\-xi\-symmetric
line--element
\begin{eqnarray}
ds^2&=&e^{-2Q(z,\bar z)}dz\,d{\bar z}\nonumber\\
&+&\frac{e^{-2G(z,\bar z)}}{a+b} \left[a(z,\bar
z)d\sigma+d\tau\right]\left[b(z,\bar z)d\sigma-d\tau\right],
\end{eqnarray}
where $\partial_\sigma$ and $\partial_\tau$ are killing vectors,
such that one is spacelike and the second is a timelike one.

The evaluation of the Newman--Penrose curvature coefficients--
Weyl complex components-- with respect to the null tetrad basis,
\begin{eqnarray}
g=2e^1\otimes e^2-2e^3\otimes e^4,
\end{eqnarray}
where
\begin{eqnarray}
&&e^1=\frac{1}{\sqrt{2}}e^{-Q}dz,\,\,\,e^3=\frac{1}{\sqrt{2}}\frac{e^{-G}}{\sqrt{a+b}}
\left(b\,d\phi-dt\right),\nonumber\\
&&e^2=\frac{1}{\sqrt{2}}e^{-Q}d{\bar z},\,\,\,e^4=\frac{1}{\sqrt{2}}\frac{e^{-G}}{\sqrt{a+b}}
\left(a\,d\phi+dt\right),
\end{eqnarray}
yields the following nonvanishing components
\begin{eqnarray}
\Psi_0&=&\frac{e^{2Q}}{a+b}\left[2\frac {\partial a}{\partial z}
\frac {\partial P}{\partial z}+ \frac {\partial^2 a}{\partial z^2}
-\frac{2}{a+b}\left(\frac{\partial a}{\partial z}\right)^2
\right],\nonumber\\
\bar{\Psi_4}&=&\frac{e^{2Q}}{a+b}\left[2\frac {\partial
b}{\partial z} \frac {\partial P}{\partial z}+ \frac {\partial^2
b}{\partial {z}^2} -\frac{2}{a+b}\left(\frac{\partial b}{\partial
z}\right)^2
\right],\nonumber\\
6\Psi_2&=&\frac{e^{2Q}}{(a+b)^2}\left[2(a+b)^2\frac {\partial^2 P}
{\partial {z}\partial {\bar z}}+5\frac {\partial
a}{\partial{z}}\frac {\partial b}{\partial{\bar z}}- \frac
{\partial a}{\partial{\bar z}}\frac {\partial b}{\partial{z}}
\label{orig} \right],\nonumber \\
\end{eqnarray}
where $P=P(z,\bar{z}):=Q-G$.
\section{Conformally flat stationary axisymmetric spacetimes; general case
$\partial a/\partial z \neq \partial b/\partial z \neq 0$
\label{ii}} If one demands the spacetime to be conformally flat,
then the Weyl tensor has to vanish, which is equivalent to fulfill
the requirements $\Psi_0=\Psi_4=\Psi_2=0$. Accordingly, one has:
\begin{eqnarray}
\Psi_0=0\,\Longrightarrow {2\frac {\partial a}{\partial z}\frac {\partial P}{\partial z}+
\frac {\partial^2 a}{\partial z^2} -\frac{2}{a+b}\left(\frac{\partial a}{\partial z}\right)^2}=0,
\label{w1}
\end{eqnarray}
and,
\begin{eqnarray}
\Psi_4=0\,\Longrightarrow {2\frac {\partial b}{\partial z}\frac {\partial P}{\partial z}+
\frac {\partial^2 b}{\partial z^2} -\frac{2}{a+b}\left(\frac{\partial b}{\partial z}\right)^2}=0.
\label{w2}
\end{eqnarray}
Subtracting equations (\ref{w1})--(\ref{w2}), one obtains
\begin{eqnarray}
\frac{\partial}{\partial
z}{\left[\frac{e^{2P}}{(a+b)^2}\left(\frac{\partial a}{\partial z}
-\frac{\partial b}{\partial z}\right)\right]}=0, \nonumber
\end{eqnarray}
therefore its integration yields
\begin{eqnarray}
{\frac{\partial a}{\partial z} -\frac{\partial b}{\partial
z}}= \bar{g}({\bar z})(a+b)^2\,e^{-2P}. \label{w3}
\end{eqnarray}
Next, dividing Eq. (\ref{w1}) by ${\partial a}/{\partial z}$ and
Eq. (\ref{w2}) by ${\partial b}/{\partial z}$, and subsequently
adding the resulting equations, one gets
\begin{eqnarray}
\frac{\partial}{\partial
z}\ln{\left[\frac{e^{4P}}{(a+b)^2}\frac{\partial a} {\partial
z}\frac{\partial b}{\partial z}\right]}=0, \nonumber
\end{eqnarray}
thus
\begin{eqnarray}
{\frac{\partial a} {\partial z}\frac{\partial b}{\partial z}}=
\bar{h}({\bar z})(a+b)^2\,e^{-4P}. \label{w6}
\end{eqnarray}
For real $\Psi_2$, one arrives at the condition
\begin{eqnarray}
\frac{\partial a}{\partial z}\frac{\partial b}{\partial \bar {z}}=\frac{\partial a}
{\partial \bar{ z}}\frac{\partial b}{\partial {z}}.
\label{w4}
\end{eqnarray}
Since the functions $a$, $b$, and $P$ are real, therefore from Eq.
(\ref{w3}) one has
\begin{eqnarray}
 g(z){\frac{\partial }{\partial
z}}(a-b)= \bar{g}(\bar z)\frac{\partial } {\partial \bar z}(a-b).
\end{eqnarray}

\noindent By introducing a new variable $z$, such that
$g(z){\frac{\partial }{\partial z}}\rightarrow {\frac{\partial }
{\partial z}} $, the above equation becomes
\begin{eqnarray}
\left({\frac{\partial }{\partial z}}-{\frac{\partial }{\partial
\bar z}}\right)(a-b)=0,
\end{eqnarray}
This transformation, allowed because of the freedom in the choice
of the variable $z$, is equivalent to set $g(z)=1$.
\noindent Hence
$a-b=F(z+\bar z)$. Introducing the real coordinates $x$ and $y$
through $z=x+iy$, denoting with dots the derivatives with respect
to $x$, and substituting $a=b+F(x)$ into Eq. (\ref{w4}), one
obtains
\begin{eqnarray}
\dot{F}\left({\frac{\partial }{\partial z}}-{\frac{\partial
}{\partial \bar z}}\right)b=0.
\end{eqnarray}
\noindent If $\dot{F}=0$, then using linear transformations of the
Killingian variables $\tau$ and $\sigma$, one can achieve $a=b$,
and consequently the metric is static.

\subsection{Metric for $\dot{F}\neq 0$, $b=b(z+\bar z)$ and $a=a(z+\bar z)$}

\noindent If now $\dot{F}\neq 0$, then $b=b(x)$ and $a=a(x)$.
Without loss of generality, equations (\ref{w3}) and (\ref{w6})
rewrite co\-rres\-pon\-din\-gly
\begin{eqnarray}
\dot{a}-\dot{b}=(a+b)^2 \,\,e^{-2P},
\label{w7}
\end{eqnarray}
and
\begin{eqnarray}
\dot{a}\dot{b}=\epsilon k^2(a+b)^2 \,\,e^{-4P}, \label{w8}
\end{eqnarray}
where  the parameter $\epsilon=\pm {1}$. Introducing new dependent
functions $X=X(x)$ and $Y=Y(x)$ on the variable $x$ according with
\begin{eqnarray}
&&{a}+{b}=2k\,Y,\,\,a=k(Y+X),\nonumber\\
&&{a}-{b}=2k\,X,\,\,b=k(Y-X),
\label{w10}
\end{eqnarray}
the Eq.~(\ref{w7}) becomes
\begin{eqnarray}
\dot{X}=\frac{dX}{dx}=2kY^2e^{-2P}\rightarrow
{dx=\frac{dX}{2kY^2}e^{2P}}, \label{w20}
\end{eqnarray}
while Eq.~(\ref{w8}) amounts to
\begin{eqnarray}
\dot{Y}^2-\dot{X}^2=4k^2\epsilon{Y}^2 e^{-4P}\label{w21}.
\end{eqnarray}
Substituting $e^{-2P}$ from Eq. (\ref{w20}) into Eq. (\ref{w21})
one obtains
\begin{eqnarray}
\dot{Y}^2-\dot{X}^2=\epsilon \frac{\dot{X}^2}{{Y}^2}\label{w23}
\end{eqnarray}
or equivalently, dividing by $\dot{X}\neq 0$, one has
\begin{eqnarray}
\left(\frac{dY}{dX}\right)^2=1+\frac{\epsilon}{Y^2}\rightarrow{\frac{dY}{dX}=\nu\,
\frac{\sqrt{Y^2+\epsilon}}{Y}},\,\nu=\pm 1, \label{d2}
\end{eqnarray}
with general integral
\begin{eqnarray}
\nu\,({X}-X_0)= \sqrt{Y^2+\epsilon}\,, \rightarrow
Y^2=({X}-X_0)^2-{\epsilon}, \label{sol2a}
\end{eqnarray}
where $X_0$ is an integration constant.

It remains to integrate the equation arising from $\Psi_2=0$,
\begin{eqnarray}
6\Psi_2&=& \frac{e^{2Q}}{(a+b)^2}[(a+b)^2
P_{,x,x}+4a_{,x}b_{,x}]\nonumber\\
&=&\frac{e^{2Q}}{Y^2}\left[2Y^2\,P_{,x,x}+(Y_{,x})^2-(X_{,x})^2\right].
\end{eqnarray}
Using instead of $x$ the new variable $X$, $\frac{d}{dx}
=\dot{X}\frac{d}{dX}$, $\dot{P}=\dot{X}P_{,X}=:\dot{X}P^{\prime}$,
taking into account Eq. (\ref{w23}), one gets
\begin{eqnarray}
6\Psi_2=
\frac{e^{2Q}}{Y^2}{\dot{X}}^2\left[2Y^2\,\left(P^{\prime\prime}+\frac{\ddot
X}{{\dot{X}}^2} P^{\prime}\right)+\frac{\epsilon}{Y^2}\right].
\end{eqnarray}
Therefore, for vanishing $\Psi_2$, one has
\begin{eqnarray}
P^{\prime\prime}+\frac{\ddot X}{{\dot{X}}^2}P^{\prime}+\frac{\epsilon}{2Y^4}=0.
\label{cf1}
\end{eqnarray}
Differentiation of  $\dot{X}$ from Eq. (\ref{w20}) with respect to
$x$, yields
\begin{eqnarray}
\ddot{X}=4k\,\dot{X}[Y\,Y^{\prime}-Y^2P^{\prime}]e^{-2P}=2{\dot{X}}^2
\left({Y^{\prime}}/{Y}- P^{\prime}\right)\label{cf}.
\end{eqnarray}
Replacing $\ddot{X}$ from the above relation into Eq. (\ref{cf1}),
one arrives at
\begin{eqnarray}
P^{\prime\prime}-2{P^{\prime}}^2+2\frac{Y^{\prime}}{Y}P^{\prime}
+\frac{\epsilon}{2Y^4}=0. \label{cf2}
\end{eqnarray}
Introducing the function $K=\exp(-2P)$, the Eq. (\ref{cf2})
becomes
\begin{eqnarray}
K^{\prime\prime}+2\frac{Y^{\prime}}{Y}K^{\prime}-\frac{\epsilon}{Y^4}K=0.
\label{cf2n}
\end{eqnarray}
Substituting in this equation $Y^{\prime}$ from Eq. (\ref{d2}),
and $Y$ from Eq. (\ref{sol2a}) in terms of $X$, considering that
$\nu^2=1$, one has
\begin{eqnarray}
&&[(X-X_0)^2-\epsilon]^2\,K^{\prime\prime}\nonumber\\
&+&2\,[(X-X_0)^2-\epsilon](X-X_0)
\,K^{\prime} -\epsilon\, K=0. \label{cf3}
\end{eqnarray}
To obtain the general solution of Eq. (\ref{cf3}), one
accomplishes the change $K(X)=M(X)[(X-X_0)^2-\epsilon]^{-1/2}$,
which yields
\begin{eqnarray}
M^{\prime\prime}=0\rightarrow{M(X)=C_0+C_1X}.
\end{eqnarray}
Therefore
\begin{eqnarray}
e^{-2P}=K(X)&=&(C_0+C_1X)[(X-X_0)^2-\epsilon]^{-1/2}\nonumber\\
&=&(C_0+C_1X)/Y. \label{cf5a}
\end{eqnarray}
In terms of the new coordinates $X$, after trivial sca\-ling and
coordinate translations: $2ky\rightarrow y,\; X-X_0\rightarrow
X,\; \sqrt{2}(\tau+kX_0\sigma)\rightarrow
T,\;\sqrt{2}k\sigma\rightarrow \Phi $, the studied metric amounts
to
\begin{eqnarray}
ds^2&=&e^{-2G( X, y)}\frac{1}{\sqrt{{ X}^2-\epsilon}}
\left[\frac{{d X}^2} {(C_0+C_1 X) ({
X}^2-\epsilon)}\right.\nonumber\\
&+&(C_0+C_1 X){d y}^2 +\left.k \left( -\epsilon d\Phi^2-2 X d\Phi
\,dT-dT^2\right) \right].\nonumber \\\label{cfg}
\end{eqnarray}
To establish whether this metric could be diagonalized or not, one
accomplishes additionally  further linear transformations in the
metric--sector
\begin{eqnarray}
{d\Sigma}^2 :=k \left( -\epsilon d\Phi^2-2 X d\Phi
\,dT-dT^2\right)
\label{cfg4}
\end{eqnarray}
 of the Killingian variables $T$ and $\Phi$ of the form
\begin{eqnarray}
&&dT=\alpha d\tau^\prime+\beta d\sigma^\prime,\nonumber\\
&&d\Phi=\gamma d\tau^\prime+\delta
d\sigma^\prime,\;\;\alpha\delta-\beta\gamma\neq 0, \label{32}
\end{eqnarray}
for real ${\alpha,\beta,\gamma}$, and ${\delta}$.
\noindent The
$g_{\tau^\prime\sigma^\prime}$ component of the metric--sector
${d\Sigma}^2$ amounts to
\begin{eqnarray}
g_{\tau^\prime\sigma^\prime}=-k\left[\epsilon \gamma\delta+{
X}(\beta\gamma+\alpha\delta)+\alpha\beta\right]
\end{eqnarray}
thus, $g_{\tau^\prime\sigma^\prime}$ may vanish is there exists a
real solution of the equations
\begin{eqnarray}
&&\beta\gamma+\alpha\delta=0,\nonumber\\
&&\alpha\beta+\epsilon \gamma\delta=0.
\end{eqnarray}
The general solution of this system is given by
\begin{eqnarray}
&&\alpha=\pm \gamma\sqrt{\epsilon}, \gamma=\gamma,\nonumber\\
&&\beta=\mp \delta\sqrt{\epsilon}, \delta=\delta,\label{t1}
\end{eqnarray}
therefore these constants are real parameters only in the ca\-se
$\epsilon=1.$ Accordingly, the metric--sector  components
$g_{\tau^\prime\tau^\prime}$ and $g_{\sigma^\prime\sigma^\prime}$
acquire the form
\begin{eqnarray}
g_{\tau^\prime\tau^\prime}=-2k\gamma^2(\epsilon\pm\sqrt{\epsilon}{
X}),\;\; g_{\sigma^\prime\sigma^\prime}=-2k
\delta^2(\epsilon\mp\sqrt{\epsilon}{ X}),
\end{eqnarray}
where the choice of the upper (lower) sign in
$g_{\tau^\prime\tau^\prime}$, has to be accompanied by the
choice of the upper (lower) sign in  $g_{\sigma^\prime\sigma^\prime}$.\\

Only for the branch of metrics with $\epsilon=1$ one can carry out
real linear transformations of the Killingian coordinates such
that the metric--sector ${d\Sigma}^2$ becomes dia\-go\-nal, which
on its turn implies the diagonal character of the
whole metric $ds^2$.\\

The case $\epsilon=-1$ deserves special attention; the
transformations (\ref{32}) are pure imaginary ones, and the
corresponding  metric tensor components become complex, fact
which is forbidden in real Einsteinian Relativity. \\

\section{Metric for Conformally Flat Locally Static
Spacetimes, $\epsilon=1$ \label{iii}} For $\epsilon=1$ --the only
case in which conformally flat stationary axisymmetric spacetimes
can be locally  diagonalized-- with an additional scaling
transformation of the form $\sqrt{2} \gamma
\tau^{\prime}\rightarrow \tau$ and $\sqrt{2} \delta
\sigma^{\prime}\rightarrow \sigma$, ($ X\rightarrow x$ ), the
corresponding metric becomes
\begin{eqnarray}
ds^2&=&\frac{e^{-2G(x,y)}}{\sqrt{x^2-1}}\left[\frac{dx^2}{(C_0+C_1x)(x^2-1)}+
(C_0+C_1x)dy^2 \right.\nonumber\\
&-&\left.(1\mp x){d\sigma}^2- (1\pm x){d\tau}^2 \right].
\label{cf9}
\end{eqnarray}
This branch corresponds to the choice $k=1$; the case $k=-1$
yields similar final results.
\noindent Considering the Killingian
metric sector
\begin{eqnarray}
d\Sigma^2:=-(1+x){d\sigma}^2- (1-x){d\tau}^2,
\end{eqnarray}
one faced out two possibilities:\\
A: ${x}<-1$, $\partial_{\tau}$ timelike Killing vector, and $\partial_{\sigma}$
spacelike Killing vector, \\
B: ${x}>1$, $\partial_{\tau}$ spacelike Killing vector, and $\partial_{\sigma}$
timelike Killing vector.\\
Case A: Introducing a new coordinate $A$,
\begin{eqnarray}
A^2=-\frac{1+x}{1-x},1+x<0,\,x=-\frac{A^2+1}{1-A^2},1-A^2>0,
\label{cf10}
\end{eqnarray}
identifying $\sigma\rightarrow \phi$, $\tau\rightarrow t $,
$C_0-C_1\rightarrow 2\alpha$, $C_0+C_1\rightarrow -2\beta$, and
$G(x,y)-\frac{1}{2} \ln{(\alpha+\beta A^2)/A}\rightarrow G(A,y)$,
one arrives at the expression
\begin{eqnarray}
ds^2=e^{-2G(A,y)}&&\left[\frac{dA^2}{(\alpha+\beta\,A^2)^2}+dy^2\right.\nonumber\\
&+&\left.\frac{A^2}{\alpha+\beta\,A^2}{d\phi}^2-
\frac{dt^2}{\alpha+\beta\,A^2} \right], \label{cf12}
\end{eqnarray}
Case B: Similarly as we treated the previous case, one introduces a new variable $A$, namely
\begin{eqnarray}
A^2=\frac{x-1}{x+1},\;x-1>0,\;x=\frac{1+A^2 }{1-A^2},\;1-A^2>0,
\label{cf10}
\end{eqnarray}
identifying $\sigma\rightarrow t $, $\tau\rightarrow \phi$,
$C_0+C_1\rightarrow 2\alpha$, $C_1-C_0\rightarrow 2\beta$, and
$G(x,y)-\frac{1}{2}\ln{(\alpha+\beta A^2)/A}\rightarrow G(A,y)$,
one arrives at the expression (\ref{cf12}).
\subsection{Metric for Conformally Flat Locally Static \\Spacetimes, $\epsilon=1$,
a $R\times BTZ$ representation} We would like to point out an
alternative representation of the above metric when a negative
cosmological constant is present, $\lambda\sim 1/l^2$.

By accomplishing in metric (\ref{cfg}), with $\epsilon=1$, $k=1$,
a SL(2,R)--transformation of the Killingian coordinates given by
\begin{eqnarray}
&&dT=\alpha dt+\beta d\phi,\nonumber\\
&&d\Phi=\gamma dt+\delta d\phi,\;\;\alpha\delta-\beta\gamma\neq 0,
\end{eqnarray}
where
\begin{eqnarray}
&&\alpha=-\frac{1}{\sqrt{2\,l}}(b_0r_{+}+a_0r_{-}),\;\;
\beta=\frac{1}{\sqrt{2}}(a_0r_{+}+b_0r_{-}),\nonumber\\
&&\gamma=\frac{1}{\sqrt{2\,l}}(b_0r_{+}-a_0r_{-}),\;\;
\delta=\frac{1}{\sqrt{2}}(a_0r_{+}-b_0r_{-}),\nonumber\\
&&r_{\pm}=\sqrt{\frac{l}{2}}\sqrt{M l\pm\sqrt{M^2\,l^2-J^2}},
\end{eqnarray}
with constants $C_{0}$ and $C_{1}$, appearing Eq.(\ref{cfg}),
fulfilling the relations
\begin{eqnarray}
C_{0}+C_{1}=-2a_{0}^2,\,\,C_{0}-C_{1}=-2b_{0}^2,
\end{eqnarray}
together with a transformation of the x--coordinate, $
X\rightarrow x$,
\begin{eqnarray}
 x=\frac{b_{0}^2(r^2-r_{-}^2)+a_0^2(r^2-r_{+}^2)
}{b_{0}^2(r^2-r_{-}^2)-a_0^2(r^2-r_{+}^2)}
\end{eqnarray}
where $l$ is related with the cosmological constant, $M$, and $J$
stand correspondingly for the global mass and the total angular
momentum of the BTZ solution, one arrives at a $R\times BTZ$
representation of the studied metric, namely
\begin{eqnarray}
ds^2&=&e^{-2G(r,y)}(dy^2+{ds^2}_{BTZ})\nonumber\\
&=&e^{-2G(r,y)}\left[dy^2+\frac{dr^2}{\frac{J^2}{4r^2}-
M+\frac{r^2}{l^2}} \right.\nonumber\\
&+&\left. r^2(d\phi-\frac{J}{2r^2}dt)^2
-(\frac{J^2}{4r^2}-M+\frac{r^2}{l^2})dt^2\right]. \label{bt}
\end{eqnarray}\\

\section{General Metric for Conformally Flat Stationary Axisymmetric Spacetimes \label{iv}}

From the metric (\ref{cfg}) when $\epsilon=-1$,  one arrives at a
new result: {\it there is no way to carry out  a
dia\-go\-na\-li\-za\-tion of  the whole metric $ds^2$, it remains
{\bf locally stationary axisymmetric}}. This conclusion
contradicts the theorem by Collinson, which asserts that " Every
conformally flat stationary axisymmetric spacetime is necessarily
static". Restoring the $\{x,y,\phi,t\}$--typing in metric
(\ref{cfg}); for po\-si\-tive $k$: $\sqrt{k}T\rightarrow{t}$,
$\sqrt{k}\Phi\rightarrow{\phi}$, while for $k$ negative,
$k=-\kappa$: $\sqrt{\kappa}T\rightarrow{\phi}$,
$\sqrt{\kappa}\Phi\rightarrow{-t}$, one arrives at the canonical
form of  conformally flat stationary axisymme\-tric spacetimes
\begin{eqnarray}
ds^2&=&\frac{e^{-2G(x,y)}}{\sqrt{x^2+1}}\left[\frac{dx^2}{(C_0+C_1x)(x^2+1)}+(C_0+C_1x)dy^2
\right.\nonumber\\
&+&\left.(x^2+1){d\phi}^2-(dt+x\,d\phi)^2 \right],
\label{cf8}
\end{eqnarray}

Since we started with a general form for the metric of stationary
axi\-sym\-me\-tric spacetimes, and arrived at the above expression
for this metric in the case of conformal flatness through the
determination of  the general solution of the zero--Weyl tensor
equations, the above metric is the more ge\-ne\-ral form for
conformally flat stationary axisymmetric spacetimes. Of course,
one can give other representations of this metric by using
coordinate transformations of the variable $x$ and $y$; for
instance representations in terms of tri\-go\-no\-me\-tric or
hyperbolic functions. \noindent The expression of the
factor-function $G(x,y)$ depends on the matter--field content,
i.e., on the energy--momentum tensor for different kinds of fields
in the Einstein equations; di\-ffe\-rent energy--tensors will give
rise to different $G(x,y)$.
\section{Conclusions}
On the light of the present results, we conclude that the
Collinson theorem is wrong. There exist branches of spacetimes
which are conformally flat and at the same time are stationary and
axisymmetric. The conformal factor of this class of metrics
depends on non--Killingian variables $x$ and $y$; its explicit
expression depend  on the sources of the Einstein equations.

\begin{acknowledgments}
The authors thank E. Ay\'on--Beato for useful discu\-ssions. This
work has been partially supported by CONACyT Grant 38495E.
\end{acknowledgments}

\end{document}